\documentclass[
  ,draft            
  ]
  {aipproc}

\layoutstyle{6x9}

\newcommand{\C}{{\cal C}}

\begin{document}

\title{Simulation of large deviation functions using population dynamics}


\author{Julien Tailleur}{
  address={SUPA, School of Physics and Astronomy, University of Edinburgh,
  Mayfield Road, Edinburgh EH9 3JZ, Scotland}
} 

\author{Vivien Lecomte}{
  address={DPMC, Universit\'e de Gen\`eve, 24, Quai Ernest Ansermet
  1211 Gen\`eve, Switzerland}
} 

\begin{abstract}
  In these notes we present a pedagogical account of the population
  dynamics methods recently introduced to simulate large deviation
  functions of dynamical observables in and out of equilibrium. After
  a brief introduction on large deviation functions and their
  {simulations}, we review the method of Giardin\`a \emph{et al.} for
  discrete time processes and that of Lecomte \emph{et al.} for the
  continuous time counterpart. Last we explain how these methods can
  be modified to handle static observables and extract information
  about intermediate times.
\end{abstract}

\maketitle

The main achievement of equilibrium statistical mechanics is probably
the simplification it offers in the study of static observables in
steady state. Indeed, the resolution of dynamical equations is
replaced by static averages and ensemble approaches. This can still be
very difficult, but {from} a conceptual point of view, the
problem is much simpler. {On the other hand}, when one is
interested in dynamical observables, like currents of particles, or in
out-of-equilibrium {situations}, {like for glassy or driven
systems}, such static ensemble approaches are not available anymore
and there is no general formalism on which one can rely.

For the last ten years, physicists have been interested in large
deviation functions mainly because they are good
candidates to extend the concept of thermodynamic potential{s} to
out-of-equilibrium situations and to dynamical observables (for a
review, see~\cite{Touchette}). Of course the mere definition of
out-of-equilibrium potentials is not useful in itself and the
challenge is thus to go beyond their construction. To do so, two
strategies can be followed. First, one can try to derive general
properties of large deviation functions. An example of success in
this direction is provided by the Fluctuation
Theorem~\cite{Evans1993,Gallavotti1995,Kurchan1998,Lebowitz1999},
which can be read as a symmetry of large deviation functions and is
one of the few general results valid out-of-equilibrium. Another
strategy is to consider specific examples and to compute the large
deviation function explicitly. For some simple yet non-trivial
interacting particle systems, exact computations have been possible
(for a review see~\cite{Derrida2007}), but one has to rely on numerics
for more generic systems. From the algorithmic point of view, two
paths can be followed. For small system sizes, exact procedures can be used
(see for instance~\cite{Seifert2008,Baiesi2008}) but as soon
as mesoscopic systems are considered, one has to rely on importance
sampling approaches. Generalising a procedure followed previously to
study rare events in chemical
reactions~\cite{TanaseNicola2003a,TanaseNicola2003b,Tailleur2006},
Kurchan and co-workers developed methods to compute large deviation
functions in dynamical systems~\cite{Tailleur2007}, and
discrete~\cite{Giardina2006} or continuous~\cite{Lecomte2007} time
Markov chains. We shall concentrate here on the statistical
mechanical aspects and review the methods available for Markov chains.

Let us consider a system {and call  $\{\C\}$ its set of
configurations}. As time goes from $0$ to a final time $t$, the system
typically jumps into successive configurations $\C_0,\C_1,\dots \C_K$
at distinct times $t_1,\dots t_K$. A \emph{dynamical} observable $Q$
is defined as a sum along the history of small contributions
$q_{\C_{k+1}\C_k}$, one for each \emph{transition} between two
successive configurations.  The simplest example of such observable is
probably a current of particle{s} $Q$ in a one-dimensional lattice
gas, which is either incremented of decremented every time a particle
jumps to the right or to the left, respectively. This contrasts with
static observables, like {the number of particles at a given
site}, which depend solely on the configuration of the system at a
given time. We will see {below}
that the algorithms used to
obtain the large deviation functions slightly differ in these two
cases. To characterise the fluctuations of the observable $Q$, the
first thing one can do is to extend the microcanonical approach to the
space of trajectories and compute the corresponding macrostate entropy
\begin{equation}
  s(q)=\lim_{t\to\infty} \frac 1 t {{\log}} P[Q(t)=q t].
\end{equation}
However one knows from usual statistical mechanics that working in the
microcanonical ensemble is often harder {than} in the canonical one and
we rather introduce a dynamical partition function and the
corresponding dynamical free energy
\begin{equation}
  Z(\beta,t)=\left \langle e^{- \beta Q(t)}\right \rangle;\qquad
  \label{eqn:partfunc}
\psi(\beta)=\lim_{t\to \infty} \frac 1 t \log Z(\beta,t).
\end{equation}
These definitions differ slightly from those used in the mathematical
literature, where one rather speaks about rate functions $-s(q)$
and cumulants generating functions $\psi(-\beta)$.

The main purpose of this {{short}} review is to explain how
one can compute  $\psi(\beta)$ using an
approach {relying on}
population dynamics. But let us first sketch why direct sampling would
be inefficient. Consider for simplicity the case where $s(q)$ has a
single maximum at $q_0$, which satisfies $s(q_0)=0$ {(for
normalisation purpose)}. Fluctuations around $Q=q_0t$ {which
occur with probabilities of order one} must typically be of order
$1/\sqrt{t}$ so that
\begin{equation}
  \label{eqn:exprare}
  P(Q=qt)\simeq e^{t s(q)}\simeq e^{\frac 12 t(q-q_0)^2 s''(q_0)}\sim 1,
\end{equation}
whence a probability of larger fluctuations exponentially small in
$t$. On the other hand, the dynamical partition function has a weight
$e^{-\beta Q}$ (with $Q\sim qt$) exponential in $t$. As a result,
there is a competition between the exponentially rare fluctuations and
their exponential weight such that for $\beta$ of order $1$ the
trajectories which dominate the average in (\ref{eqn:partfunc}) are
exponentially rare. {A more quantitative way to rephrase this can be
read in the Legendre relation between $s(q)$ and $\psi(\beta)$:
  $\psi(\beta)=\max_q [s(q)-\beta q].$ The maximum is realised for a
  value $q^*$
which dominates the average in \eqref{eqn:partfunc}. It thus differs
from $q_0$ -- which {maximises} only $s(q)$ -- by a factor independent of
$t$. From \eqref{eqn:exprare} we see that the corresponding
trajectories indeed have a probability exponentially small in $t$. To
have a good sampling over $N$ unbiased simulations, $N$ should thus be
of order $e^t$, an impracticable requirement. Direct sampling is thus
{a} hopeless strategy to observe large deviations.}

\section{Discrete time}
We present in this section the method first introduced by Giardin\`a,
Kurchan and Peliti~\cite{Giardina2006} to simulate cumulant generating
functions in discrete time Markov chains. In this case, the dynamics
is {defined} by the transition \emph{probabilities}
$U(\C\rightarrow\C')=U_{\C'\C}$ between configurations {and the
corresponding} the master equation reads
\begin{equation} \label{eq:master_eq_discrete_time}
  P(\C,t)=\sum_{\C'} U_{\C\C'} P(\C',t-1).
\end{equation}
Conservation of probability enforces the matrix $U$ to be stochastic,
i.e. for all $\C'$,
  $\sum_{\C}U_{\C\C'}=1$.
Starting from a fixed configuration $\C_0$ the explicit solution
of~\eqref{eq:master_eq_discrete_time} is
\begin{equation}
  P(\C,t)=\sum_{\C_1\ldots\C_{t-1}}  U_{\C\C_{t-1}} U_{\C_{t-1}\C_{t-2}} \ldots U_{\C_1\C_{0}}= \left[U^t\right]_{\C\C_0}.
             \label{eqn:explicit_sol_tdisc}
	      \label{eqn:explicit_sol_tdisc_power}
\end{equation}
The dynamical observable $Q$ can be written as a sum over
configuration changes $Q(t)=q_{\C_t\C_{t-1}}+\ldots+q_{\C_1\C_0}$. To
compute the dynamical free energy $\psi(\beta)$, we first rewrite the
dynamical partition function as
\begin{eqnarray} \label{eqn:exp_s_A_explicit}
Z(\beta,t)=  \left\langle e^{-\beta Q{(t)}}\right\rangle=
  \sum_{\C_1\ldots\C_{t}} 
   U_{\C_t\C_{t-1}}e^{-\beta\,q_{\C_t\C_{t-1}}} \ldots 
   U_{\C_1\C_{0}}  e^{-\beta\,q_{\C_{1}\C_0}}= 
  \sum_\C \left[U_\beta^t\right]_{\C\C_0},
\end{eqnarray}
where we have introduced the matrix
  $[U_\beta]_{\C\C'} = U_{\C\C'} e^{-\beta\,q_{\C\C'}}$.
Let us note that $\psi({\beta})$ is given by the log of the largest
eigenvalue of $U_\beta$.
A possible strategy, used for instance
in~\cite{Bodineau2005,Seifert2008,Baiesi2008}, is thus to compute numerically this
eigenvalue. The matrix $U_\beta$ is however exponentially large in the
system size, {which limits this strategy to small systems}. The
main advantage of this method is to yield a numerical approximation of
an exact expression - as pointed out in \cite{Baiesi2008} - as opposed
to our approach, efficient for large systems, but relying on
importance sampling. Note that the `exact' approach can be used to
check {the validity} of the importance sampling approach for small
system sizes, before going to larger ones, as was actually done for
the simulations presented in~\cite{Lecomte2007}.

Comparison of expressions~{(\ref{eqn:explicit_sol_tdisc})
and~(\ref{eqn:exp_s_A_explicit})} leads one to think that $\psi({\beta})$
could be obtained from a new dynamics, induced by $U_\beta$. However,
the matrix $U_\beta$ is not stochastic, as in general
\begin{equation}
  Y_{\C'}\equiv\sum_{\C} [U_{\beta}]_{\C\C'}= \sum_{\C} U_{\C\C'}  e^{-\beta\,q_{\C\C'}} \neq 1
\end{equation}
and we should not understand~\eqref{eqn:exp_s_A_explicit} as a stochastic
process with conserved probability but as a population dynamics with
branching and death, where the population size is not
constant. To do so, let us define
\begin{equation}
 U'_{\C \C'} = \frac{[U_{\beta}]_{\C\C'}}{Y_{\C'}}=\frac{U_{\C\C'}}{Y_{\C'}}  e^{-\beta\,q_{\C\C'}}.
\end{equation}
The matrix $U'$ is stochastic and~\eqref{eqn:exp_s_A_explicit}
now writes
\begin{eqnarray} \label{eqn:exp_s_A_Y_C}
  \left\langle e^{-\beta Q{(t)}}\right\rangle=
  \sum_{\C_1\ldots\C_{t}} 
   U'_{\C_t\C_{t-1}} Y_{\C_{t-1}} \ldots 
   U'_{\C_1\C_{0}}  Y_{\C_0}.
\end{eqnarray}
This expression is now closer
{to}~\eqref{eqn:explicit_sol_tdisc} as $U'$ is stochastic, and
can be interpreted as follows: $N$ agents evolve with the stochastic
dynamics defined by $U'$ and are replicated with a rate $Y_\C$ when
they are in configuration $\C$. This interpretation is possible as
$Y_\C$ \emph{depends solely on the initial configuration~$\C$} and can
thus be interpreted as a configuration-dependent reproduction
rate. This was less apparent in~\eqref{eqn:exp_s_A_explicit}, where
factors $e^{-\beta\, q_{\C\C'}}$ depend on both initial and final
configurations.

\bigskip

This interpretation as a population dynamics can be implemented using
a diffusion Monte Carlo algorithm. Let us consider an ensemble of
$N_0$ agents ($N_0\gg 1$) evolving in the configuration space
$\{\C\}$. At each time step {$\tau\rightarrow \tau+1$},
\begin{enumerate}
\item[(1)] Each agent evolves according to the $\beta$-modified dynamics $U'_{\C\C'}$, \label{itm:stoch_evol}

\item[(2)] Each agent in configuration $\C$ is replicated/killed with
probability $Y_\C$, $i.e.$ is replaced by $y$ copies, where

\begin{equation} 
  y=\left\{\begin{tabular}{cc} $\lfloor Y_\C\rfloor +1$& \mbox{with
    probability} $Y_\C-\lfloor Y_\C\rfloor$ \\ $\lfloor Y_\C\rfloor$ &
    \mbox{with probability} $1-(Y_\C-\lfloor Y_\C\rfloor)$ \\
    \end{tabular}\right.
\end{equation}
Concretely, the agent is replaced by $y$ copies of itself, so that the
  population size is increased by $y-1$ (decreased by $1$ if
  $y=0$). \label{itm:clonage}
\end{enumerate}
Let us show that the size of the population at time $t$ yields the
large deviation function. \emph{For a given history}, {the number
$N(\C,\tau)$ of copies in configuration $\C$ at intermediate time
$\tau$ satisfies}
  $N(\C_\tau,\tau) = 
  U'_{\C_\tau\C_{\tau-1}} Y_{\C_{\tau-1}}  N(\C_{\tau-1},\tau-1)$,
so that for the whole history
\begin{eqnarray}
  { N}(\C_t,t) &= 
  U'_{\C_t\C_{t-1}} Y_{\C_{t-1}} \ldots 
  U'_{\C_1\C_{0}}  Y_{\C_0} \: { N}(\C_0,0).
\end{eqnarray}
Consequently, we see {from~\eqref{eqn:exp_s_A_Y_C}} that the
population size ${ N}(t) = \sum_{\C} { N}(\C,t)$ behaves as
${ N}(t)/N_0=\langle e^{-\beta\,Q{(t)}} \rangle \sim e^{t\psi(\beta)}$.  As
 usual in importance sampling approaches, the ensemble average
 $\langle . \rangle$ has been replaced by an average over a finite
 number of simulations.  Whereas in principle correct, this approach
 is however impracticable {since} the population size varies
 exponentially in time and we thus add a third step to the previous
 algorithm:
\begin{enumerate}
\item[(3)] After the cloning step, the population is rescaled by a
  factor {$X_\tau$} to its initial size $N_0$, by uniformly pruning/replicating the
  agents.
  \label{itm:rescaling}
\end{enumerate}
At each time step $\tau$, the rescaling factor is given by
  $X_\tau=\frac{{ N}(\tau-1)}{{ N}(\tau)}$
so that
  $X_{t}\ldots X_0 = \frac{{ N}_0}{{ N}(t)}$
and finally
\begin{equation}
  \psi(\beta) = - \lim_{t\to\infty} \frac{1}{t}  {\log} \langle X_{t}\ldots X_0 \rangle.
\end{equation}

\section{Continuous time dynamics}

For a continuous time dynamics defined by rates $W(\C\to\C')$ the
master equation reads
\begin{equation} \label{eqn:mait}
  \partial_t P(\C,t) = \sum_{\C'\neq\C} 
  W(\C'\to\C) P(\C',t)\ - r(\C) P(\C,t),
\end{equation}
where
$r(\C)=\sum_{\C'} W(\C\to\C')$
is the escape rate from configuration $\C$.

There {are} many different ways of deriving the algorithm
presented in the previous section and we shall {follow} here a
derivation of the continuous {time} algorithm slightly different {from}
the one {we} introduced in~\cite{Lecomte2007}.
The formal solution of \eqref{eqn:mait} reads
\begin{eqnarray} 
  \label{eqn:sol_explicit} 
  &P(\C,t) = \sum_{K\geq 0} 
  \sum_{\C_1\ldots\C_{K-1}}
  \int_{t_0}^t dt_K \int_{t_0}^{t_K} dt_{K-1}  
  \ldots \int_{t_0}^{t_2} dt_1\\
&\rho(t_K|\C_{K-1},t_{K-1}) \cdots \rho(t_{1}|\C_0,t_0) e^{-(t-t_K)r(\C)} \nonumber
    \frac{W(\C_0\rightarrow\C_1)}{r(\C_0)}\ldots
  \frac{W(\C_{K-1}\rightarrow\C)}{r(\C_{K-1})},\nonumber
\end{eqnarray}
where
  $\rho(t_k|\C_{k-1},t_{k-1})=r(\C_{k-1}) \exp[-(t_k-t_{k-1})r(\C_{k-1})]$ 
represents the probability distribution of the time intervals between
jumps. The sum over $K$ corresponds to all the possible numbers of
jumps between 0 and $t$, the sum over the $\C_k$'s to the different
configurations which can be {visited. The integrals over $t_k$
{account} for all the possible times at which jumps occur}. Last, the
ratio $\frac{W(\C_{k-1}\rightarrow\C_k)}{r(\C_{k-1})}$ gives the
\emph{probability} that the system goes {to} configuration $\C_k$, when
it quits  configuration $\C_{k-1}$. Multiplying the second line of
\eqref{eqn:sol_explicit} by $e^{-\beta Q}$ yields an explicit formula
for $Z(\beta,t)$:
\begin{eqnarray} 
  \label{eqn:sol_explicit_Z} 
  Z(\beta,t) =& \sum_{K\geq 0} 
  \sum_{\C_1\ldots\C_{K-1},\C}
  \int_{t_0}^t dt_K \int_{t_0}^{t_K} dt_{K-1}  
  \ldots \int_{t_0}^{t_2} dt_1\\
  &  \rho(t_K|\C_{K-1},t_{K-1}) \cdots \rho(t_{1}|\C_0,t_0) e^{-(t-t_K)r(\C)} \nonumber\\
  &  \frac{W(\C_0\rightarrow\C_1)}{r(\C_0)}e^{-\beta q_{\C_1 \C_0}}\ldots
  \frac{W(\C_{K-1}\rightarrow\C)}{r(\C_{K-1})}e^{-\beta {q_{\C \C_{K-1}}}}.\nonumber
\end{eqnarray}
Further introducing the biased rates
$W_\beta(\C\to\C')=W(\C\to\C') e^{-\beta q_{\C'\C}}$,
 the corresponding escape rates $r_\beta$ and time distributions
between two jumps $\rho_\beta$, \eqref{eqn:sol_explicit_Z} can be
rewritten {(after some algebra)} as
\begin{eqnarray}
  \label{eqn:sol_explicit_Z2} 
  Z(\beta,t) =& \sum_{K\geq 0} 
  \sum_{\C_1\ldots\C_{K-1},\C}
  \int_{t_0}^t dt_K \int_{t_0}^{t_K} dt_{K-1}  
  \ldots \int_{t_0}^{t_2} dt_1\\
  &  \rho_\beta(t_K|\C_{K-1},t_{K-1}) \cdots \rho_\beta(t_{1}|\C_0,t_0) e^{-(t-t_K)r_\beta(\C)} \nonumber\\
  &  {Y}(\C_0)^{t_1-t_0} \frac{W_\beta(\C_0\rightarrow\C_1)}{r_\beta(\C_0)}\ldots
	 { Y}(\C_{K-1})^{t_K-t_{K-1}}  \frac{W_\beta(\C_{K-1}\rightarrow\C)}{r_\beta(\C_{K-1})} {Y}(\C)^{t-t_K},\nonumber
\end{eqnarray}
where
{${Y}(\C_k)=e^{r_\beta(\C_k)-r(\C_k)}$}.
{$Z(\beta,t)$ is thus a weighted sum over all possible
trajectories generated by the biased rates $W_\beta$, where the
weights are given by the factors ${Y}(\C_k)^{t_{k+1}-t_k}$}. A first
idea which can come to mind is to simply evolve the population  
with the {rates} $W_\beta$, without cloning and to simply average $
e^{\int_0^t {d\tau} (r_\beta(\C(\tau))-r(\C(\tau)))}$ over these
trajectories, as was proposed in~\cite{Imparato2007}. This however
fails as soon as $t$ is large, for the same reason as {the one}
described in the introduction: {the weight is exponentially large
in $t$ and large fluctuations of the exponent are exponentially
rare}. One thus has to use a biased sampling to compute the average
\eqref{eqn:sol_explicit_Z2}. {Following the philosophy of the
``Go with the Winner methods''~\cite{Grassberger2002}, the general
idea is to stochastically replace a trajectory with weight $\mathcal
W$ by `$\mathcal W$' trajectories with weight $1$, so that trajectories
which high rates are favoured whereas those with small weights are not
investigated.}

If the re-weighting procedure is made systematic, every time an agent
$c_\alpha$ changes of {configuration} at time $t^\alpha$, one gets the
following algorithm:
\begin{enumerate}
\item[(0)] The time is set to $t^\alpha$.
\item[(1)] $c_\alpha$ jumps from its configuration $\C$
  to another configuration $\C'$ with probability $W_\beta(\C\to\C')/r_\beta(\C)$.
\item[(2)] The time interval $\Delta t$ until the next jump of
 $c_\alpha$ is chosen from the Poisson law $\rho_\beta$ of parameter
  $r_\beta(\C')$.
\item[(3)] The agent $c_\alpha$ is either cloned or pruned with a rate ${\cal Y}(\C')= e^{\Delta t(r_\beta(\C')-r(\C'))}$
  \begin{itemize}
  \item[a)] One computes $y=\lfloor {\cal Y}(\C')+\epsilon\rfloor$ where $\epsilon$ is uniformly
    distributed on $[0,1]$.
  \item[b)] If $y=0$, the copy $c_\alpha$ is erased.
  \item[c)] If $y>1$, we make $y-1$ new copies of $c_\alpha$.
  \end{itemize}
\item[(4)] If $y=0$, one agent $c_\beta\neq c_\alpha$ is chosen at
  random and copied, while if $y>1$, $y-1$ agents are chosen uniformly
  among the $N+y-1$ agents and erased. We store the rescaling factor $X=\frac N {N+y-1}$.
\end{enumerate}

To reconstruct the dynamical free energy, we keep track of all $X$
  factors
\begin{equation}
  \label{eq:main_result} 
  \frac{1}{t}{\log} \langle X_1 \ldots
  X_{\tau}\rangle =\frac{1}{t} {\log} \left\langle e^{-\beta Q(t)} \right\rangle \sim
  -\psi(\beta) \qquad \mbox{as} \qquad {t\to\infty}
\end{equation}
Once again the step (4) ensures constant population.

\section{The case of static observables}
\label{sec:static}
The methods presented above {only} apply for dynamical
observables, which can be decomposed as sums of individual
contributions over each configuration {change}. One could also be
interested in averages of static observables along the histories
$O=\int_0^td\tau \:o(\tau)=\sum_k (t_{k+1}-t_k)o(\C_k)$. In this case,
the above procedure simplifies and the algorithm is identical apart
from two points.
\begin{itemize}
\item First, there is no bias in the rate. The agents are 
evolved with the unmodified Markov rates $W(\C\to\C')$.
\item Second, the cloning rate is simply given by
  $e^{-\beta(t_{k+1}-t_k) o(\C_{k})}$.
\end{itemize}
This can best be seen in formula \eqref{eqn:sol_explicit_Z} by
replacing the weight $e^{-\beta q_{\C_{k+1}\C_k}}$ which depends on
the configurations before and after the jump by
$e^{-\beta(t_{k+1}-t_k) o(\C_{k})}$ which depends solely on the
configuration $\C_k$ and can thus be seen as a constant cloning rate
for the whole time the system spends in the configuration $\C_k$.

\section{Intermediate times}

As pointed out in~\cite{Giardina2006}, the configurations probed along
the simulation are representative of the typical ones at 
\emph{final time} $t$ in the evolution, rather than at intermediate times
($0\ll {\tau} \ll t$).  In particular, the weighted average value $\langle
O(t)\rangle_\beta$ of a static observable $O(\C(t))$ at final time $t$ is
obtained in the algorithm by computing the average of $O$ among the
agents at the end of the simulation.

In general, the value of $\langle O({\tau})\rangle_\beta$ at
intermediate times $0\ll {\tau} \ll t$ differs from the one at final
time, yet $\langle O({\tau})\rangle_\beta$ is of particular interest
since it is representative of configurations visited during most of
the weighted evolution (see~\cite{sglassy} for examples).
{A way to compute $\langle O({\tau})\rangle_\beta$ numerically
can be read in its formal expression:
\begin{eqnarray}
  \langle O(\tau) \rangle_\beta =& \sum_{K\geq 0} \sum_{\C_1\ldots\C_{K-1},\C}
  \int_{t_0}^t dt_K \int_{t_0}^{t_K} dt_{K-1}  
  \ldots \int_{t_0}^{t_2} dt_1 O(\tau)\\
  &  \rho_\beta(t_K|\C_{K-1},t_{K-1}) \cdots \rho_\beta(t_{1}|\C_0,t_0) e^{-(t-t_K)r_\beta(\C)} \nonumber\\
  &  {Y}(\C_0)^{t_1-t_0} \frac{W_\beta(\C_0\rightarrow\C_1)}{r_\beta(\C_0)}\ldots
	 { Y}(\C_{K-1})^{t_K-t_{K-1}}  \frac{W_\beta(\C_{K-1}\rightarrow\C)}{r_\beta(\C_{K-1})} {Y}(\C)^{t-t_K}\nonumber
\end{eqnarray}}
One simply runs the same algorithm as before, which generates
the bias on trajectories, except that whenever an agent arrives at a
time $t_k$ such that $t_{k-1}\leq {\tau} <t_k$, the corresponding
value $O(\C_{k-1})$ 
is
attached to the agent. Then, each time an agent is cloned, the
corresponding value of $O$ is copied accordingly. At the end of the
simulation, $\langle O({\tau})\rangle_\beta$ is obtained from the
average of the values of $O(\tau)$ attached to the surviving clones.
Of course, thanks to the cloning process between ${\tau}$ and $t$,
this average differs from the one we could have done at the
intermediate time ${\tau}$ in the simulation.

The same kind of scheme also applies to compute the weighted average
of any observable $O$ depending on the whole history of the system and
the crucial {step} is to copy the value of the observable when cloning
events occur.
Note in particular that the determination of $\langle
O({\tau})\rangle_\beta$ can be quite noisy since only a few instances
of $O(\tau)$ have survived at time $t$. {In the long time limit
one may gain similar information by studying $\frac 1t \big\langle
\int_0^t d{\tau} O(\C({\tau}))\big\rangle_\beta$ \cite{sglassy} which
is a less noisy dynamical observable.}

\section{Discussion}

These numerical methods have been applied successfully in many
different situations but it is important to keep in mind their
limitations. First, as pointed out in~\cite{Harris2008} convergence
problems are met when the evolution operator is gapless. This can for
instance happen in systems where the configuration space is unbounded,
as in the Zero Range Process but will however not be a problem as long
as the configuration space remains finite.

\begin{figure}[htbp]
  \centering
  \includegraphics[width=.48\columnwidth]{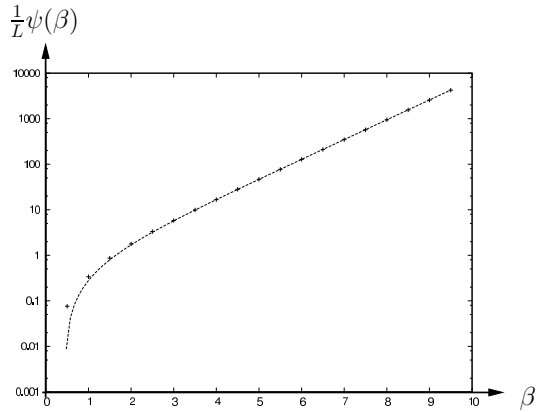}
  \caption{Large deviation function $\psi(\beta)/L$ (in log scale) for
    the total current of particles in the SSEP
    at density $\rho=1/2$ ($L=400$ sites). The points (+) are the result of the
    continuous-time numerical algorithm. The line is the analytic
    result~\eqref{eq:psiQ_SEP_sgrand} valid for very large deviations
    $|\beta|\gg 1$. }
  \label{fig:SEPbigbeta}
\end{figure}

The other important limitation is the finiteness of the population of
agents, which can be problematic for very large deviations yielding
large cloning rates. For all applications considered
in~\cite{Lecomte2007}, we thus ensured that the cloning factor would
never grow larger than few percent of the total population size and
agreement with theory - when at hand - was very good. 
As an example, one can compare (figure~\ref{fig:SEPbigbeta}) for the simple symmetric
exclusion process (SSEP) the numerical
result in the regime of very large deviations with the analytical
result~\cite{TheseV}
\begin{equation} 
\label{eq:psiQ_SEP_sgrand}
  \frac 1L\psi(\beta)\:=\:2 \cosh \beta \,\frac{\sin\pi \rho}{\pi}-2\rho(1-\rho)-
  2\frac{\sin^2(\pi\rho)}{\pi^2}+{\mathcal O}(e^{-|\beta|})
\end{equation}
valid for $|\beta|\gg 1$, for the total current of particles in the system.
Agreement is very good although the values of $\beta$ correspond to very large deviations.
For large
cloning rates, it may also be necessary to modify step (4) of the
continuous time algorithm so that agents are not pruned uniformly but
according to the weight they carry since their last change of
configuration $\propto e^{(t-t_k)[r_\beta(\C_k)-r(\C_k)]}$ but in our
simulations we never ran into this problem. In this worst case
scenario, the efficiency of the continuous time  {implementation} would fall back to
that of the discrete time where at every {step} the whole population is
re-sampled.




\begin{theacknowledgments}
We thank R.J. Harris, A. {R\'akos} and H. Touchette for many
useful discussions and P. Hurtado for suggesting this review. JT
acknowledges funding from EPSRC grants EP/030173.
{VL was supported by the Swiss
FNS, under MaNEP and division II.}

\end{theacknowledgments}

\bibliographystyle{aipproc}   

\end{document}